\documentclass[draft]{agujournal2019}
\usepackage{url} 
\usepackage{lineno}
\usepackage[inline]{trackchanges} 
\usepackage{soul}
\usepackage{xspace}

\draftfalse

\journalname{Geophysical Research Letters}

\begin{document}

%
%


\title{Shallow cumulus convection over land in cloud-resolving simulations with a coupled ray tracer}

%
%




\authors{M. A. Veerman\affil{1}, B. J. H. van Stratum\affil{1}, and C. C. van Heerwaarden\affil{1}}

\affiliation{1}{Meteorology and Air Quality Group, Wageningen University, The Netherlands}

\affiliation{1}{PO BOX 47, 6700 AA, Wageningen, The Netherlands}




\correspondingauthor{Menno Veerman}{menno.veerman@wur.nl}

\begin{keypoints}
\item We present GPU-accelerated cloud-resolving simulations with a coupled ray tracer.
\item Simulated cloud statistics show low sensitivity to the level of convergence of the ray tracer.
\item Strong coupling between surface irradiance patterns and cloud evolution underlines need for coupled 3D radiative transfer.
\end{keypoints}

\begin{abstract}
We present simulations of cumulus convection over land with shortwave radiation computed by a Monte Carlo ray tracer coupled to the flow solver. Ray tracing delivers very realistic in-cloud heating rates and global horizontal irradiance fields. The necessary performance of the ray tracer has been enabled by the raw power of GPU computing and from techniques for accelerating data lookup and ray tracer convergence. We used a case study over a grassland in the Netherlands to compare simulations with a coupled ray tracer to those with a conventional two-stream solver, and to test ray tracer convergence. We demonstrate that the simulated cloud evolution is insensitive across a wide range of samples sizes in the ray tracer.
Furthermore, simulations with a coupled ray tracer produce surface irradiance patterns that resemble observations and that strongly feed back to the evolution of clouds via locally enhanced surface heat fluxes.
\end{abstract}

\section*{Plain Language Summary}
Clouds absorb and reflect solar radiation and create spatial patterns at the land surface of cloud shadows interspersed with sunny regions.
These patterns are currently largely simplified in most weather models because radiation computations cost a lot of computer power.
We have developed a fast and realistic numerical radiation model that runs on a modern computer graphics card.
By doing so, we can do simulations that produce radiation patterns that closely resemble reality and study how these patterns affect clouds.
This enables us to improve our understanding of the complex interactions between clouds, solar radiation, and the Earth’s surface. 

\section{Introduction}

Solar radiation enters the atmosphere following the direct beam of the sun, and can be absorbed or scattered in any direction by gas molecules, aerosols, cloud droplets, and the surface. 
This 3D nature of radiation affects the spatial structure of atmospheric radiative heating and global horizontal irradiance.
Clouds intercept more radiation at their sides as the solar zenith angle increases, which enhances the size of cloud shadows \cite<side illumination;>{Hogan2013}, but may also result in more diffuse irradiance \cite<side escape;>{Hogan2013}.
Additionally, clouds can intercept radiation by the surface or by neighbouring clouds \cite<entrapment;>{Hogan2019}.
These solar radiation -- cloud interactions create complex surface patterns with cloud shadows and regions where global horizontal irradiance exceeds clear-sky radiation.

Ideally, one would capture these 3D effects in the radiation computations of cloud-resolving simulations.
Monte Carlo ray tracing is one of the methods that can do so \cite<e.g.>{Mayer2009, Villefranque2019}, and can produce patterns in global horizontal irradiance that closely resemble field observations \cite{Gristey2020}.
However, ray tracing was long deemed to be too computationally costly to be applied coupled to 3D atmospheric models with a sufficiently large domain size and detail level \cite{Cahalan2005}.
The actual status quo in weather and climate models is the use of relatively efficient two-stream methods that solve radiation in the vertical direction \cite{Meador1980, Cahalan2005}.
These approaches are computationally affordable, but lack the aforementioned 3D radiative effects.

The development of the TenStream solver \cite{Jakub2015}, which reduces the direction of photon propagation to a limited set, allowed for detailed studies into 3D radiative effects in a coupled cloud-resolving model. 
These studies \cite{Jakub2017,Veerman2020} revealed clear impacts of 3D radiative effects on cloud development, mainly driven by global horizontal irradiance patterns that strongly deviate from those in simulations with two-stream solvers. 

The fast increase in computational power and memory of Graphics Processing Units (GPUs) has dramatically improved the perspective of coupling ray tracers to atmospheric simulations.
Ray tracing is one of the core applications of modern GPUs, primarily in the gaming and movie industry, and GPU computing has proven itself as a promising technique for cloud-resolving simulations \cite{Schalkwijk2015,Heerwaarden2017,Esparza2022} in recent years.
In this study, we leverage the computing capabilities of modern GPUs to better understand how 3D radiative effects impact atmospheric simulations.
We developed a GPU-accelerated ray tracer and integrated this into GPU-resident cloud-resolving model MicroHH \cite{Heerwaarden2017}.
Using a case study of cumulus convection over land, we then compare simulations with coupled ray tracing to those with a conventional two-stream solver to address the following questions:
\begin{enumerate}
    \item What is the required ray tracer convergence and how sensitive are clouds to it?
    \item How do interactions between clouds, land surface, and radiation compare between simulations with ray tracer and two-stream solvers?
\end{enumerate}

\section{Methods}
\subsection{The case study}
Our case study is designed based on observations of the Cabauw station in the Netherlands, taken at August 15, 2016 and documented in \citeA{Tijhuis2022}.
Initial conditions, large scale forcings and background profiles for radiative transfer computations are constructed from ERA5 reanalysis \cite{Hersbach2020}.
The Cabauw station is part of the Baseline Surface Radiation Network \cite{Knap2022} and 1 Hz solar irradiance observations are available \cite{obsdata1,obsdata2}.
In short, our case is a day with shallow cumulus developing around 8 UTC (10 LT) and moderate wind speeds of approximately 5 m s$^{-1}$ in the convective boundary layer. 
Cloud base and top are around 1300 m and 2000 m, respectively, and cloud cover fluctuates between 0.3 and 0.7 during the afternoon. 
The surface is well-watered grassland and we use an albedo of 0.22 for all wavelengths.

\subsection{The cloud-resolving model}
We simulated the case with MicroHH \cite{Heerwaarden2017} following \citeA{Tijhuis2022}, but with a domain size of 38.4 $\times$ 38.4 $\times$ 4 km${^3}$ and a grid spacing of 50 m in the horizontal and 20 m in the vertical.
The radiative transfer scheme is called every 60 seconds.
Solar radiative transfer can be computed using either the two-stream method of RTE+RRTMGP \cite<Radiative Transfer for Energetics + RRTM for General circulations model applications - Parallel;>{Pincus2019} or with Monte Carlo ray tracing (Section \ref{sect:mm-rt}).
Longwave radiative transfer computations are done in 1D and neglect scattering.
We use the sets of 112 shortwave and 128 longwave spectral quadrature points available in RTE+RRTMGP.
MicroHH uses a surface model that closely follows \citeA{Balsamo2009} and parameterizes warm two-moment microphysics following \citeA{Seifert2001}.

\subsection{The ray tracer}\label{sect:mm-rt}
The newly developed ray tracer extends the open-source C++/CUDA implementation of RTE+RRTMGP and has been designed for running on a GPU.
Modern GPUs are very suitable for ray tracing, because they can execute many rays simultaneously, have high memory bandwidth, and can store entire 3D fields into memory, which avoids the need for distributed computing.
While the latter gives our GPU implementation its main performance edge compared to parallel CPU solvers, the memory of a single GPU also puts an upper limit to the problem size.

The ray tracer solves the radiative transfer equation with a Monte Carlo approach, i.e. tracing many rays through the domain until radiative fluxes and heating rates have sufficiently converged.
It requires 3D fields of atmospheric optical properties as input: total extinction coefficient $k_\mathrm{ext}$, cloud extinction coefficient $k_\mathrm{ext, cloud}$, single scattering albedo $\omega_0$ and asymmetry parameter $g$.
While the default RRTGMP solves these optical properties for a subset of the domain for all spectral quadrature points simultaneously, we built a version optimized for 3D radiative transfer that solves the full domain per quadrature point. Ray tracing is then performed independently for each quadrature point.

Rays are initialized at the domain top (TOD) following a scrambled Sobol sequence.
First the total incoming flux and the diffuse fraction at the TOD are computed from a clear-sky 1D atmospheric profile spanning from the surface to the TOA using a two-stream solver.
Pre-computing the top-of-domain flux allows the ray tracer to be confined to a limited vertical extent, which reduces computational complexity.
However, by doing so we likely underestimate the top-of-domain flux by a few W m$^{-2}$ by neglecting the effect of the underlying cloud field.
Depending on the diffuse fraction, the initial direction of each photon is either along the sun beam or randomly generated, assuming isotropic downwelling diffuse radiation at the TOD. 
Rays are then propagated through the atmosphere using a null-collision approach with an acceleration grid similar to \citeA{Villefranque2019} but without recursion:
a 3D grid with a coarser resolution than the simulation is constructed by selecting the highest extinction coefficient within each grid cell as null-collision coefficient.
Ray tracing is then performed on the acceleration grid, while local optical properties are retrieved from the simulation grid.

We use the Rayleigh phase function for scattering of gases and the Henyey-Greenstein phase function without delta-scaling for scattering by cloud droplets. 
The Henyey-Greenstein phase function does not fully capture the forward peak of the Mie phase function and is therefore too diffusive, but this may improve convergence and thus reduce computational costs compared to using the actual Mie phase function.
Within clouds, the relative probability of being scattered by a cloud droplet as opposed to a gas molecule is given by $k_\mathrm{ext,cloud}/k_\mathrm{ext}$.
Absorption is handled by using weights, which accelerates convergence of radiative fluxes and heating rates.
Each ray is assigned an initial weight of unity and its weight is reduced at each scattering or null-collision event by multiplying it with the local probability that the ray will not be absorbed \cite{Iwabuchi2006}. 
The mean weight of all rays hitting the surface or leaving the domain top is then used to compute surface and TOD radiative fluxes. 
Additionally, we store the weight of the incoming ray multiplied with the local absorption probability at each scattering or null-collision event to compute 3D radiative heating rates.
We rescale the weights following \citeA{Iwabuchi2006}: whenever the weight $w$ is lower than 0.5, we draw a random number $\rho$ between 0 and 1. The ray is terminated if $\rho > w$, else the weight is set to 1. 
The surface is assumed to be Lambertian, e.g. scattering isotropically. 
The local surface albedo is used to reduce the weight after scattering.

\subsection{The experiments}
Our experiment consists of 9 simulations (Table \ref{tab:runs}). $\texttt{rt}$ and $\texttt{2s}$ are the two main simulations, using the ray tracer and two-stream solver for solar radiative fluxes, respectively. 
\texttt{rt-s032}, \texttt{rt-s064}, \texttt{rt-s128} are similar to \texttt{rt}, but have fewer samples per pixel ($N_\mathrm{sample}$; the number of rays entering the top of the domain per grid column per spectral quadrature point).
These simulations are used to investigate how the convergence of the ray tracer affects the simulation. 
Additionally, we take 10 cloud fields from the \texttt{rt} and \texttt{2s} simulations to study the convergence of the ray tracer offline: 
Radiative fluxes and heating rates are computed for each field with $N_\mathrm{sample} = \left( 16, 32, 64, 128, 256, 512, 1024 \right)$ and compared to a benchmark computation with $N_\mathrm{sample} = 16384$.

The second set of simulations (Table \ref{tab:runs}) aims to determine the contributions of surface irradiance patterns, mean surface irradiance, and radiative heating rate patterns on the cloud evolution in \texttt{rt} and \texttt{2s}. 
In \texttt{rt-hom} and \texttt{2s-hom} the downwelling and upwelling surface solar radiative fluxes are horizontally homogenized, so that only the horizontally averaged solar irradiance drives the surface heat fluxes.
In \texttt{rt-hom-hr} and \texttt{2s-hom-hr} the solar radiative heating rates are horizontally homogenized.

\section{Results}
\subsection{First impression}
\begin{figure}
\includegraphics[width=\linewidth]{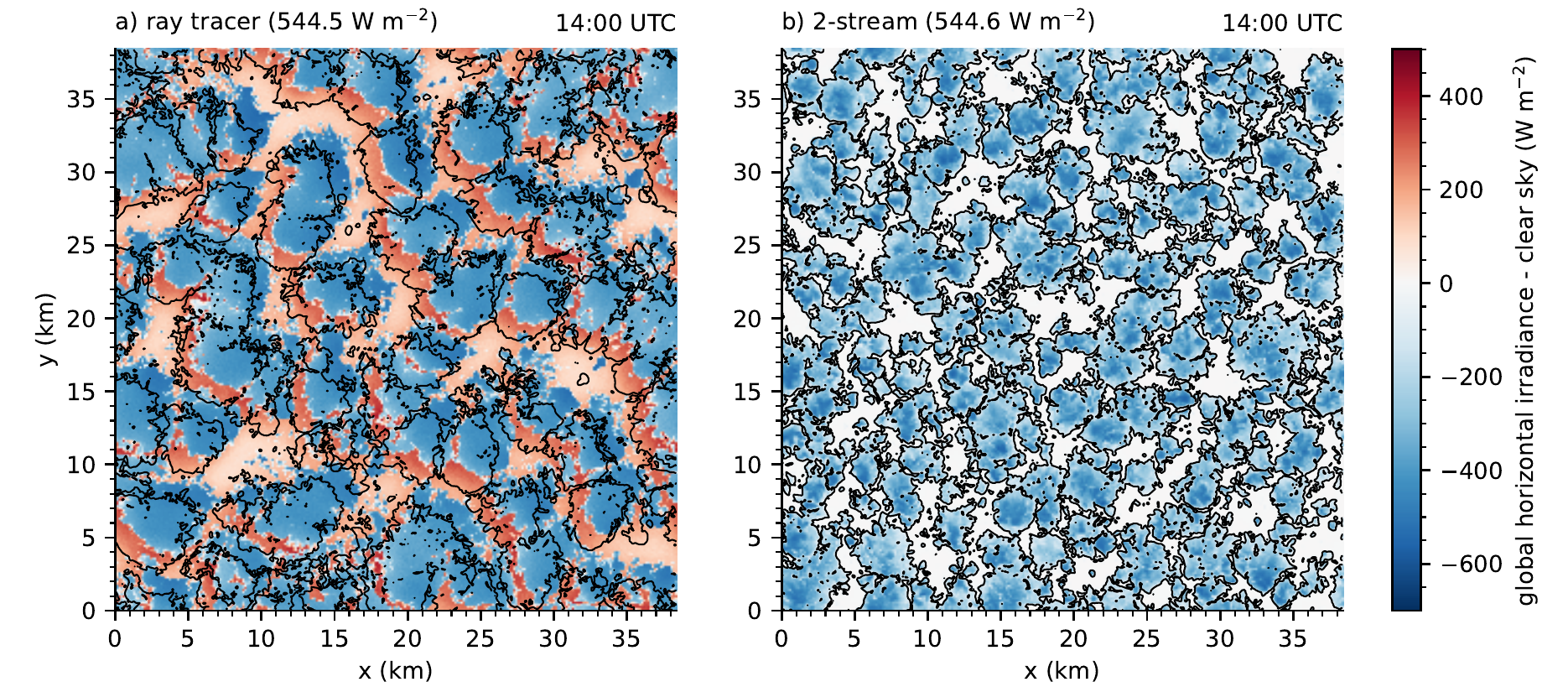}
\caption{\textbf{Simulated global horizontal irradiance.} Global horizontal solar irradiance at 14 UTC plotted relative to clear-sky radiation for a) the simulation with ray tracer, and b) the simulation with two-stream solver. Black lines are contours of liquid water path (LWP $>$ 0 kg m$\mathrm{^{-2}}$). The sun is approximately in the south-west (azimuth angle: 228.1$^\circ$, zenith angle: 46.4$^\circ$).}
\label{fig:cross}
\end{figure}

The global horizontal irradiance (GHI) fields at 14 UTC (Figure \ref{fig:cross}) display clear differences between the reference ray tracing \texttt{rt} and two-stream \texttt{2s} simulations.
The GHI field in \texttt{rt} shows the horizontally shifted and elongated cloud shadows and the GHI enhancements of clear-sky (non-shadow) regions, which are both absent in \texttt{2s}.
Furthermore, the liquid water path contours show fewer but larger clouds in \texttt{rt} compared to \texttt{2s}.
Before interpreting the differences between the \texttt{rt} and \texttt{2s} simulations, we first need to verify whether ray tracer convergence is sufficient for the chosen number of samples per pixel ($N_\mathrm{sample}$ = 256).

\subsection{Convergence of ray tracer and associated computational costs}

\begin{figure}
\includegraphics[width=\linewidth]{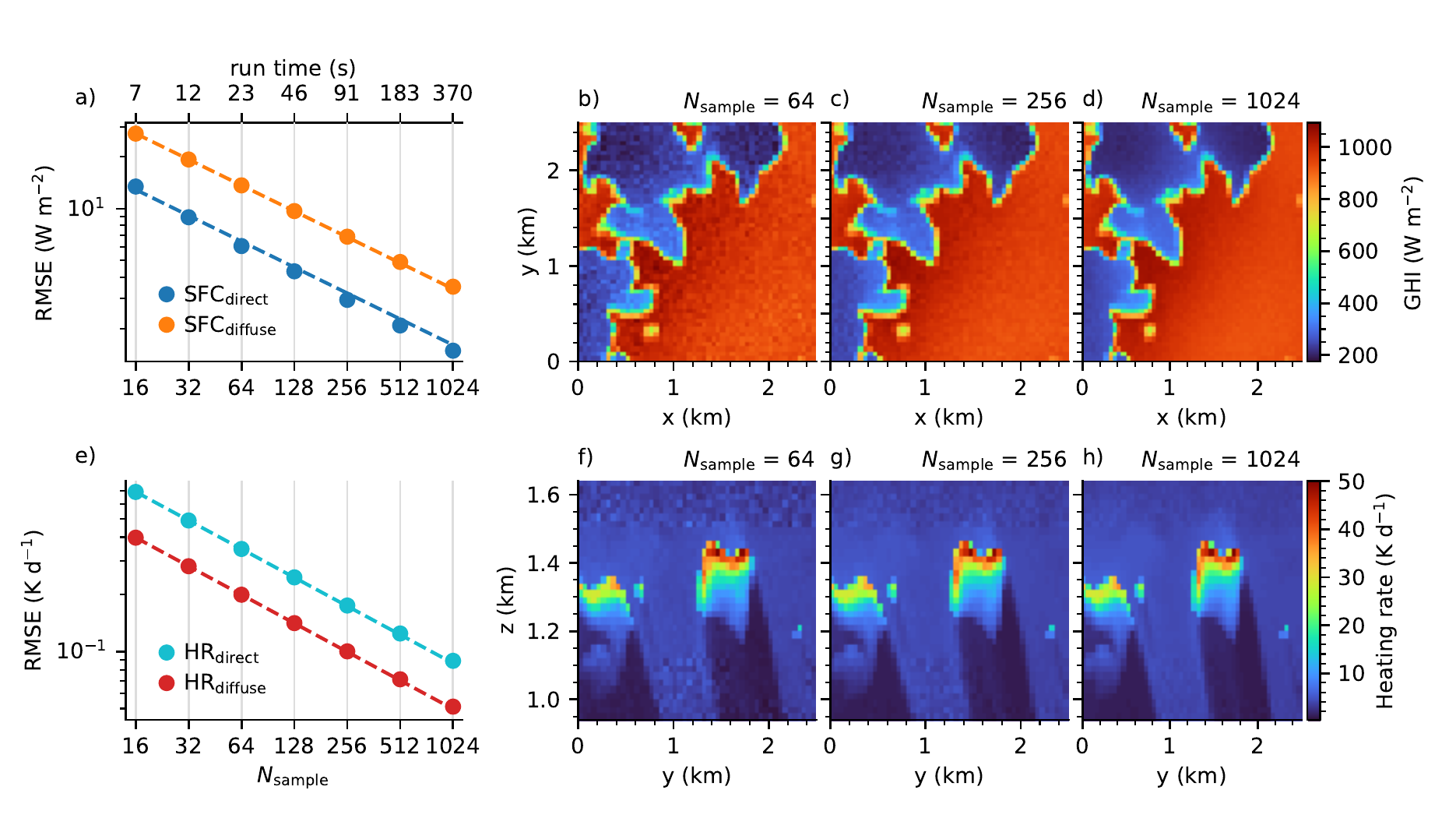}
\caption{\textbf{Convergence of ray tracer results}.
Root mean square errors (RMSE) of surface direct (horizontal) and diffuse irradiance ($\mathit{SFC}\mathrm{_{direct}}$, $\mathit{SFC}\mathrm{_{diffuse}}$) (a), and direct and diffuse radiative heating rates ($\mathit{HR}\mathrm{_{direct}}$, $\mathit{HR}\mathrm{_{diffuse}}$) (e) as function of $N_\mathrm{sample}$, computed over 10 cloud fields with respect to $N_\mathrm{sample}$ = 16384. Horizontal cross-sections of global horizontal irradiance (b-d) and vertical cross-sections of radiative heating rates (f-h) computed with 64 (b, f), 256 (c, g) and 1024 (d, h) samples per pixel for an identical cloud field. Dashed lines in a) and e) show curve fits of the form: $\mathit{RMSE}=aN_\mathrm{sample}^{-1/2}$.}
\label{fig:conv_offline}
\end{figure}

The convergence of the global horizontal irradiances (GHI; Figure \ref{fig:conv_offline}b-d) and radiative heating rates (HR; Figure \ref{fig:conv_offline}f-h) predicted by the ray tracer visibly improves from 64 to 1024 samples per pixel.
The large-scale patterns in GHI and HR (e.g. cloud shadows, the cloud enhancements, cloud-top absorption) are already visible with $N_\mathrm{sample} = 64$, but the small-scale noise in the cross-sections reduces as $N_\mathrm{sample}$ increases. 
As expected, the decrease in root mean square error (RMSE) with $N_\mathrm{sample}$ is approximately proportional to $N_\mathrm{sample}^{-1/2}$, whereas the run time increases approximately linearly (Figure \ref{fig:conv_offline}a, e).
With $N_\mathrm{sample} = 256$, as used in the main ray tracing simulations (\texttt{rt}, \texttt{rt-hom}, \texttt{rt-hom-hr}), the ray tracing for all 112 $g$-points takes approximately 90 seconds for one cloud field, but errors are not negligible:  
the RMSE of the diffuse irradiance is 6.88 W m$^\mathrm{-2}$ and the RMSE of the direct heating rates is 0.17 K d$^\mathrm{-1}$.

\begin{figure}
\includegraphics[width=.935\linewidth]{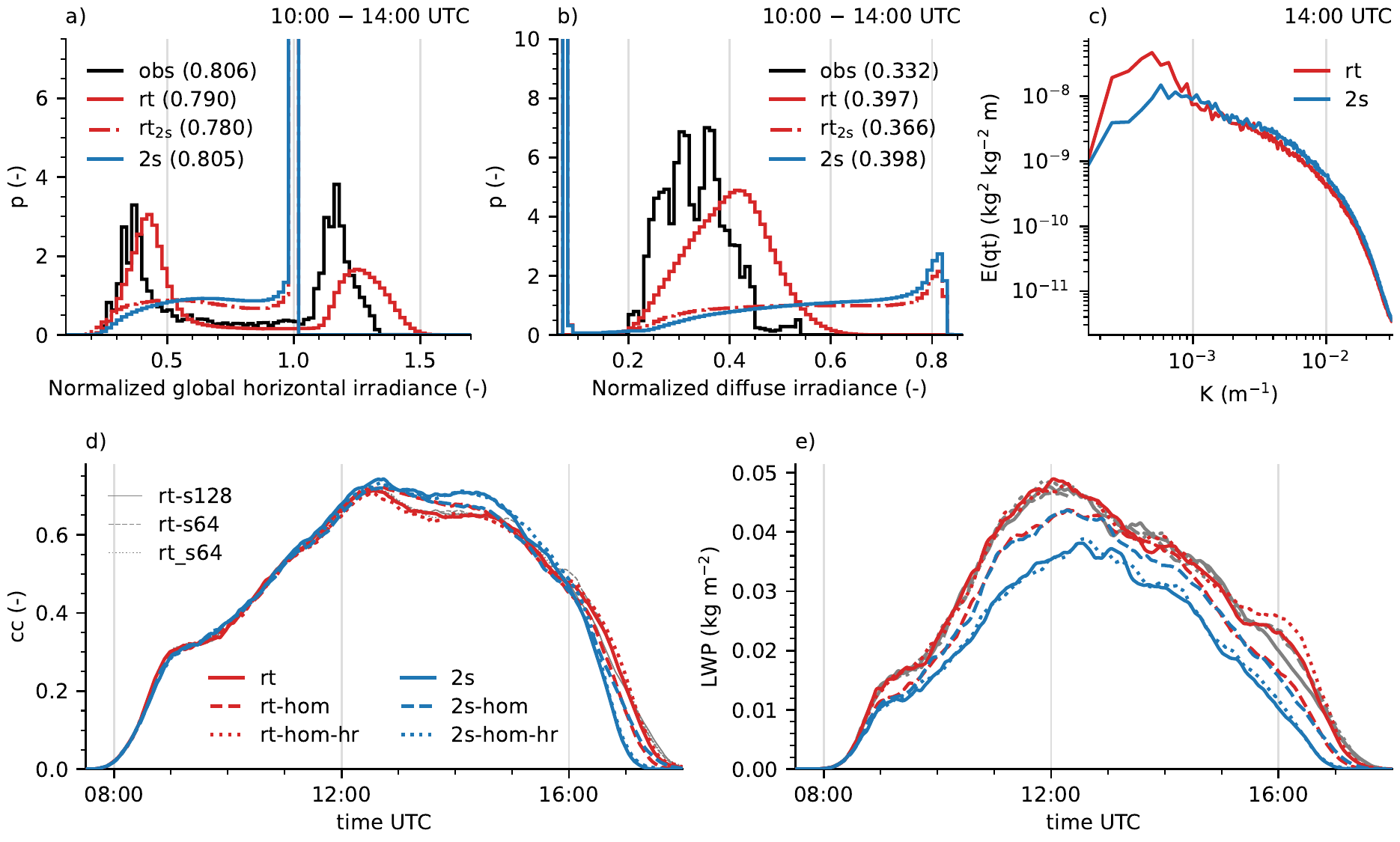}
\caption{\textbf{Surface irradiance and cloud statistics.} Probability density functions (mean values between brackets) of global horizontal irradiance (a) and diffuse irradiance (b) between 10 and 14 UTC, normalised by clear-sky irradiance, for the \texttt{rt} and \texttt{2s} simulations, for 1 Hz irradiance observations, and for offline two-stream computations based on the \texttt{rt} simulation (\texttt{rt$_\mathrm{2s}$}). Spectra of specific humidity $\mathrm{q_t}$ (c) for the \texttt{rt} and \texttt{2s} simulations at 14 UTC, averaged over the middle of the cloud layer (1500 $-$ 1700 m). Time series of cloud cover (d) and mean liquid water path (e) for the \texttt{rt} and \texttt{2s} simulations and the simulations with horizontally homogenized surface irradiances (\texttt{-hom}) or heating rates (\texttt{-hom-hr}).}
\label{fig:irr_ts_hist}
\end{figure}

However, the effect of the ray tracer convergence on the simulated behavior of the atmosphere is negligible:
differences in cloud cover and liquid water path between \texttt{rt}, \texttt{rt-s128}, \texttt{rt-s64}, and \texttt{rt-s32} are small and within the statistical noise (Figure \ref{fig:irr_ts_hist}d, e).

\subsection{Interactions between 3D radiation and clouds}

\subsubsection{Surface irradiance}
Online 3D radiative transfer affects global horizontal irradiance directly due to 3D cloud radiative effects \cite<e.g. entrapment, cloud-side illumination;>{Hogan2013,Hogan2019}, and indirectly via changes in cloud development.
In \texttt{rt}, the global horizontal irradiance (GHI) is lower than in \texttt{2s} (Figure \ref{fig:irr_ts_hist}a). 
Comparing \texttt{2s} to \texttt{rt$_\mathrm{_{2s}}$}, which is an offline two-stream computation based on \texttt{rt} fields, suggests that this lower GHI in \texttt{rt} is primarily caused by cloud field differences. 
Ray tracing actually enhances the GHI compared to the offline two-stream computations, due to an increase in diffuse irradiance (DIF; Figure \ref{fig:irr_ts_hist}b), despite a small decrease in direct irradiance as a result of cloud-side illumination (not shown). 

Differences in the spatial distribution of the GHI and DIF between \texttt{rt} and \texttt{2s} are more pronounced than the domain-mean differences.
The GHI PDF of \texttt{rt} (Figure \ref{fig:irr_ts_hist}a) shows the characteristic bimodal distribution that is also found in irradiance observations below broken clouds but generally missed by two-stream approximations (see e.g. \citeA{Gristey2020,Tijhuis2022}): 
a peak at low GHI corresponding to cloud shadows receiving mostly diffuse radiation and a peak at high GHI corresponding to clear-sky regions receiving both direct and diffuse radiation. 
Similarly, the DIF PDF of \texttt{rt} has an approximately gaussian shape that resembles observations, in contrast to the DIF PDFs of \texttt{2s} and \texttt{rt$_\mathrm{2s}$} (Figure \ref{fig:irr_ts_hist}b).
The GHI PDF of \texttt{rt} is shifted towards larger values compared to observations, whereas the mean GHI is lower in \texttt{rt}.
This shift is presumably due to the higher DIF in \texttt{rt} compared to observations, possibly caused by differences between the simulated and actual clouds fields.

However, differences in the PDFs may also be attributed to uncertainties in the radiative transfer computations or the observations. 
The simulations do not include aerosols and therefore likely overestimate GHI and underestimate DIF \cite{Gristey2022}.
Furthermore,the modelled direct radiation is defined as all non-scattered radiation, whereas the observed direct radiation is all radiation from a solid angle of 5.0$^\circ$ centered around the sun \cite{sensor2001}.
Radiation scattered in almost forward direction would therefore be counted as direct radiation in observations, but as diffuse in the ray tracer.
However, the difference between both definitions is likely small, because the Henyey-Greenstein phase function is too diffusive compared to the Mie phase function.
The response time of the thermopile radiation sensors (95\% of the signal after 7s) \cite{sensor2001,sensor2004}, which smooths out the actual cloud-induced radiation signal, may also contribute to the differences between the observed and modelled PDFs. 

\subsubsection{Cloud development}
In \texttt{rt}, the cloud cover is lower, but the liquid water path (LWP) is much larger than in \texttt{2s} (Figures \ref{fig:irr_ts_hist}d, e). 
This suggests that 3D radiative transfer may result in thicker clouds, as earlier shown by 
\citeA{Veerman2020} for a different case of shallow cumulus near Cabauw. 
The spectra of specific humidity within the cloud layer have significantly more energy at larger spatial scales in \texttt{rt}, which is consistent with the cloud sizes in Figure \ref{fig:cross}.
This indicates that in \texttt{rt}, the turbulence field within the cloud layer is dominated by larger structures, suggesting that individual clouds are wider. 

The simulations with horizontally homogenized surface irradiances (\texttt{rt-hom}, \texttt{2s-hom}) have similar cloud covers and LWPs during most of the day, especially compared to the differences between \texttt{rt} and \texttt{2s}.
In contrast, horizontally homogenizing the radiative heating rates (\texttt{rt-hom-hr}, \texttt{2s-hom-hr}) only has a small impact on cloud development. 
Again, this illustrates that differences in cloud development between \texttt{rt} and \texttt{2s} are mainly driven by 3D radiative effects on GHI distributions.
However, it is know that 3D longwave radiative transfer affects cloud development by altering radiative heating within clouds\cite{Klinger2017}, so the role of 3D radiative heating rates found here may be too weak because we solve longwave radiative transfer in 1D 

\begin{figure}
\includegraphics[width=0.66\linewidth]{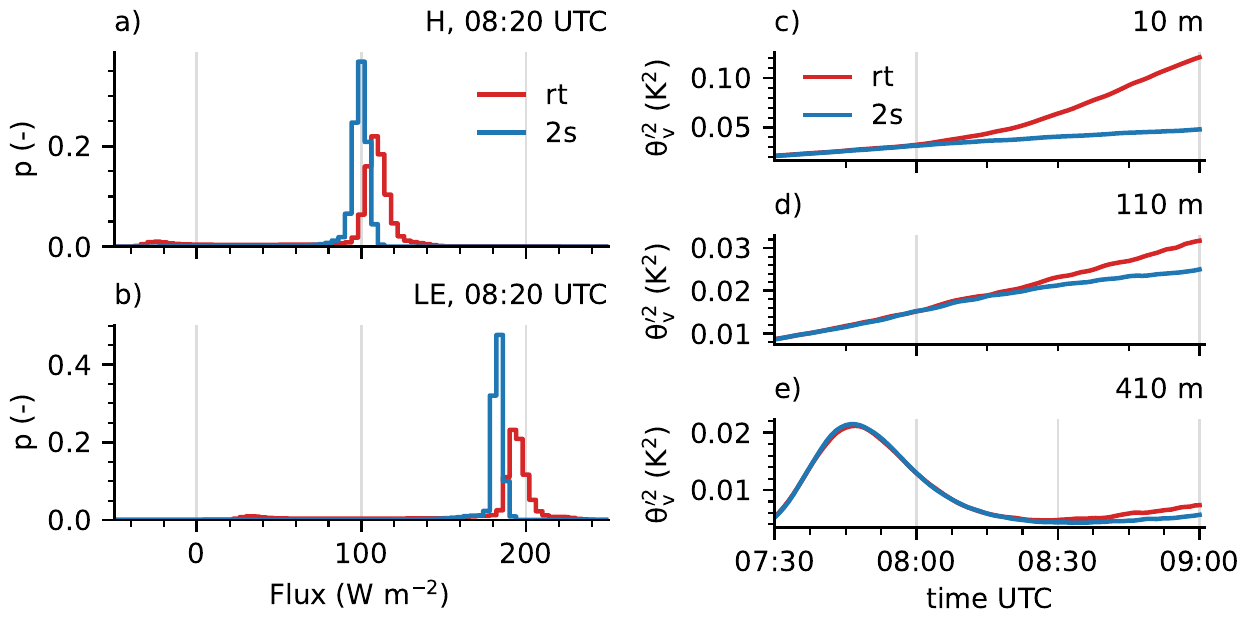}
\caption{Probability density functions of a) sensible (H) and b) latent (LE) heat fluxes, and time series of the virtual potential temperature variance $\theta_v^{\prime2}$ at heights of 10 m, 110 m and 410 m.}
\label{fig:seb_thv}
\end{figure}

To unravel the chain of events that make GHI patterns impact convection, we focus on the surface heat fluxes and the variance of the virtual potential temperature $\theta^{\prime 2}_v$ around 8:20 UTC, before the cloud statistics start diverging (Figures \ref{fig:seb_thv}a, b).
The mean sensible and latent heat fluxes in \texttt{rt} and \texttt{2s} are very similar at this time, but their spatial distributions already differ substantially.
\texttt{rt} has patches with higher surface heat fluxes than \texttt{2s}, which can generate stronger plumes \cite{Patton2005, vanHeerwaarden2014}.
This changes the buoyancy field near the surface, represented by $\theta_v^{\prime2}$, which then propagates upward towards the cloud layer (Figures \ref{fig:seb_thv}c-e):
$\theta_v^{\prime2} $ starts diverging near the surface (10 m) just after 8 UTC, near the top of the surface layer (110 m) around 8:15 UTC and at 410 m around 8:30 UTC, which approximately coincides with the moment of diverging cloud statistics (cloud base is about 550 m around 8:30 UTC).

\section{Conclusions}
This work demonstrates cloud-resolving simulation over land with a coupled ray tracer. GPU computing permits studies to solar radiation -- cloud interactions at a high detail level (50 m horizontal grid spacing) and horizontal domain size (38.4 $\times$ 38.4 km$^2$).

Our work corroborates earlier findings that the primary route of 3D radiative effects to influence cloud formation and evolution is via the land surface, and adds fidelity to this finding by delivering simulations with global horizontal irradiance patterns that closely resemble observations.
Furthermore, as our test case is an actual observed case with a moderate background wind, the presented modifications are not constrained to idealized studies under calm conditions.

Compared to other techniques to incorporate 3D radiative effects in cloud-resolving simulations, ray tracing makes few assumptions and simplifications, while delivering realistic in-cloud heating patterns and surface irradiance fields.
In addition, the technical implementation of the ray tracer is of low complexity and memory usage, opening up the prospect of addressing larger problems than the one in this study.
Therefore, cloud-resolving simulation with coupled ray tracing has large potential to become an important tool in unraveling the earth's radiation balance.

 \begin{table}\label{tbl:exp}
 \caption{\textbf{Runs performed in experiment.} $N_\mathrm{sample}$ shows samples per horizontal grid cell. Homogenization describes horizontal averaging. \textit{Total cost} is the run time (on a NVIDIA A100 with 40 GiB memory), of the simulation, measured between 13:00 UTC and 13:15 UTC and \textit{Radiation cost} the average time spend in the radiative transfer scheme.}\label{tab:runs}
 \centering
 \begin{tabular}{l c r c r r}
 \hline
  run  & Radiation solver & $N_\mathrm{sample}$ & Homogenization & Total cost (s) & Radiation cost (s)\\ 
 \hline
   \texttt{rt} & ray tracer & 256 & none & 1709.2 & 103.6\\
   \texttt{2s} & two-stream & n/a & none & 292.1 & 5.6 \\
   \hline
   \texttt{rt-s128} & ray tracer & 128 & none & 1160.1 & 65.0\\
   \texttt{rt-s064} & ray tracer & 64  & none & 881.2 & 45.4\\
   \texttt{rt-s032} & ray tracer & 32  & none & 748.0 & 36.1 \\
   \hline
   \texttt{rt-hom} & ray tracer & 256 & surface & 1703.0 & 106.3\\
   \texttt{2s-hom} & two-stream & n/a & surface & 292.7 & 5.6\\
   \texttt{rt-hom-hr} & ray tracer & 256 & heating rates & 1696.4 & 103.3\\
   \texttt{2s-hom-hr} & two-stream & n/a & heating rates & 306.3 & 5.6 \\
 \hline
 \end{tabular}
 \end{table}
 
\section{Open Research}
The CUDA version of RTE+RRTMGP, including ray tracer, and the used version of MicroHH, including setups of all simulations, are available at \url{https://zenodo.org/record/7099927}.
Model data used for the figures in this manuscript are available at \url{https://doi.org/10.5281/zenodo.7100850}. Observational data is available at \url{https://zenodo.org/record/7093164} and \url{https://zenodo.org/record/7092058}. 
\acknowledgments
All authors acknowledge funding from the Dutch Research Council (NWO) (grant: VI.Vidi.192.068).
B.v.S. acknowledges funding from NWO (grant 184.034.015). The simulations were carried out on the Dutch national e-infrastructure with the support of SURF Cooperative.
The authors acknowledge help from the ESiWACE2 service call in improving the GPU performance of the radiation code.
The authors thank Robert Pincus for many useful discussions on 3D radiation.


\end{document}